# Mapping of Low-Frequency Raman Modes in CVD-Grown Transition Metal Dichalcogenides: Layer Number, Stacking Orientation and Resonant Effects

Maria O'Brien[1,2†], Niall McEvoy[1,2†*], Damien Hanlon[2,3], Toby Hallam[2,3], Jonathan N. Coleman[2,3] and Georg S. Duesberg[1,2*]

[1]School of Chemistry, Trinity College Dublin, Dublin 2, Ireland

[2]Centre for Research on Adaptive Nanostructures and Nanodevices (CRANN) and Advanced Materials and BioEngineering Research (AMBER) Centre, Trinity College Dublin, Dublin 2, Ireland

[3]School of Physics, Trinity College Dublin, Dublin 2, Ireland

†These authors contributed equally.

## Abstract

Layered inorganic materials, such as the transition metal dichalcogenides (TMDs), have attracted much attention due to their exceptional electronic and optical properties. Reliable synthesis and characterization of these materials must be developed if these properties are to be exploited. Herein, we present low-frequency Raman analysis of $MoS_2$, $MoSe_2$, $WSe_2$ and $WS_2$ grown by chemical vapour deposition (CVD). Raman spectra are acquired over large areas allowing changes in the position and intensity of the shear and layer-breathing modes to be visualized in maps. This allows detailed characterization of mono- and few-layered TMDs which is complementary to well-established (high-frequency) Raman and photoluminescence spectroscopy. This study presents a major stepping stone in fundamental understanding of layered materials as mapping the low-frequency modes allows the quality, symmetry, stacking configuration and layer number of 2D materials to be probed over large areas. In addition, we report on anomalous resonance effects in the low-frequency region of the $WS_2$ Raman spectrum.

**Introduction**

Transition metal dichalcogenides (TMDs), such as $MoS_2$ and $MoSe_2$, have recently attracted significant attention from both industry and academia due to their wide range of fascinating properties[1-3]. Unlike graphene, these materials possess a sizable bandgap and many reports indicate that they could be suitable as active layers in logic electronics and optoelectronics[3-6] and as constituents in a variety of energy related applications[7-10]. High-quality monolayer flakes of TMDs have previously been obtained via mechanical exfoliation[1-3]; however, this method is serendipitous and suffers from low-throughput. Chemical[11] and liquid-phase exfoliation[12-15] have greatly improved the prospect of scalability, however, the crystals produced by these methods typically have relatively small lateral dimensions rendering them ill-suited for many electronic applications. Large-scale TMD films have been obtained by sulfurization of metal oxide[16, 17] or metal films[18-20], but, the thus derived films are typically polycrystalline. Recently, there have been significant advances using chemical vapour deposition (CVD)[21-28] to produce large-area, high-quality crystals, which could facilitate the realization of industry-relevant devices. In the case of each of these aforementioned synthesis routes it is imperative that the composition, quality and thickness of the materials produced is assessed before they can be considered for use in applications. Techniques such as atomic force microscopy and transmission electron microscopy are useful in the characterization of layer number and crystalline quality, respectively, but suffer from low sample throughput and laborious sample preparation.

Raman spectroscopy is a widely used technique in materials science and can be used to study molecular vibrations in 2D materials, which can reveal a wealth of information about material properties in a fast and non-destructive manner. In the case of graphene, Raman spectroscopy can be used to investigate the number and relative orientation of individual atomic layers, and can provide information on defect levels, strain and doping[29]. Recent studies have shown that analogous information can be obtained for TMD samples, with each TMD having a characteristic spectrum. $MoS_2$ is the most heavily studied TMD to date and numerous reports on its Raman characteristics, and their dependence on layer number, have emerged. The most commonly reported Raman characteristics are those corresponding to reasonably large energy shifts, such as the in-plane $E^1_{2g}$ and the out-of-plane $A_{1g}$ mode, which are observed at ~385 and ~405 $cm^{-1}$, respectively. Additional modes can be observed in the low-frequency (<50 $cm^{-1}$) region of the Raman spectrum of TMDs, known as the shear modes (SMs) and layer-breathing modes (LBMs) and recent reports have demonstrated the practicality of studying these modes[30]. These low-frequency modes occur due to relative motions of the planes themselves, either perpendicular or parallel to the atomic layers, and can prove useful in the characterization of 2D materials.

Herein, we present a systematic study of the low-frequency Raman peak positions and intensities of CVD-grown TMDs, including $MoS_2$, $MoSe_2$, $WSe_2$ and $WS_2$. These peaks were mapped out over large areas in regions consisting of crystals with different layer thickness, as are often found in CVD-grown samples, demonstrating the feasibility of using low-frequency Raman mapping for assessing layer number in TMD crystals. The same areas were also characterized using standard (high-frequency) Raman spectroscopy and photoluminescence (PL) spectroscopy. We identify different stacking configurations in $MoSe_2$ and $WSe_2$ by detailed analysis of Raman spectra and maps. Lastly, a newly observed resonant Raman mode, related to the *LA(M)* mode, has also been identified in the low-frequency region of the $WS_2$ Raman spectrum.

**Results and Discussion**

TMDs can exist in 3 polytypes, depending on the co-ordination of chalcogen atoms around the metal atoms, and the stacking order of the layers. The first, 1T, is a metallic crystal with octahedral co-ordination that has recently been artificially synthesized for device applications[31]. However, since this polytype is metastable and not found in nature[32], we will not discuss it here. The more common 2H and 3R polytypes are semiconducting, with trigonal prismatic coordination, with similar properties but differing stacking orders of metallic and chalcogen atoms. For example, 2H has a stacking order of AbA BaB AbA BaB, where capital letters indicate chalcogen atoms and lower-case letters indicate metal atoms, while 3R has a typical stacking order of AbA BcB CaC AbA, or the inverted AbA CaC BcB AbA[32]. The layers can also adopt a mixture of these stacking configurations, whereby, for example in a 3L sample, layers 1-2 obey 2H stacking, and layers 2-3 obey 3R stacking[33]. This means that a 3L 2H-3R sample could have the stacking configuration AbA BaB CbC, or AbA BaB AcA. The properties of 2H and 3R TMDs have been reported to be almost identical[32], with little observable change in the high-frequency region of the Raman spectrum. However, recent reports indicate slight differences in band structures and absorption spectra between the two stacking types[34, 35]. This shows that further investigation into the identification and properties of these stacking configurations is important both for fundamental studies of these materials and for future studies in the emerging field of van der Waals heterostructures[36] where the stacking order of two dissimilar layers could change the electronic and optical properties[37, 38] of artificial[39] or grown[40] heterostacks. In this study, we refer to 2H stacked crystals unless explicitly stated otherwise.

The Raman spectra of 2H and 3R semiconducting TMDs generally display two main characteristic vibrational modes. These are the $E'/E_g/E^1_{2g}$ and $A'_1/A_{1g}$ first-order modes at the

Brillouin zone centre, shown in Figure 1, that result from the in-plane and out-of-plane vibrations, respectively, of metal (M) and chalcogen (X) atoms[41-43]. Different peak labels are used for different layer numbers due to the changing symmetry of the point group from $D_{3h}$ (odd layer number) to $D_{3d}$ (even layer number) to $D_{6h}$ (bulk). These Raman active modes have been shown to shift in position with number of layers[26, 44-46], allowing mono- and few-layer crystals to be identified. For example, in the case of $MoS_2$, as the layer number increases, interlayer van der Waals (vdW) forces suppress atomic vibrations meaning higher force constants are observed[44]. This means that the out-of-plane $A'_1/A_{1g}$ mode becomes blue-shifted at higher layer numbers (~2 cm$^{-1}$ from monolayer to bilayer), as the vibrations of this mode are more strongly affected by vdW forces between the layers. The in-plane $E'/E_g/E^1_{2g}$ mode in contrast shows a red shift as layer number increases (~2 cm$^{-1}$ from monolayer to bilayer). This is attributed to structural changes in the material or to an increase in long-range Coulombic interlayer interactions affecting the atomic vibrations[30, 44]. However, for the transition metal diselenides, such as $MoSe_2$ and $WSe_2$, these changes in frequency for different layer numbers are not as dramatic (e.g. a shift of ~1 cm$^{-1}$ in the $A_{1g}$ peak from 2 to 3L $MoSe_2$) [45, 47], and may be below the instrumental spectral resolution of standard equipment. Furthermore, crystallite size[48], doping and strain have been shown to significantly alter the Raman spectra of TMDs. Previous reports have shown a red shift and broadening of the $A'_1/A_{1g}$ peak in $MoS_2$ with n-doping[49] and a blue shift and enhancement of the $A'_1/A_{1g}$ peak with p-doping[50]. The Raman spectrum of $MoS_2$ is also highly sensitive to strain with the application of uniaxial strain resulting in the degeneracy of the $E'/E_g/E^1_{2g}$ mode being lifted[51], whereas the introduction of localized wrinkles and folds has been shown to cause a red shift of both $A'_1/A_{1g}$ and $E'/E_g/E^1_{2g}$ modes[52]. Given the large number of factors that can affect the primary peaks in the Raman spectra of TMDs, an alternative method for the clear assessment of TMD layer numbers using Raman spectroscopy is desirable.

Investigation of the low-frequency SM and LBM has been suggested as a universal method of layer number (N) determination in TMD materials[30], due to the fact that the layer-breathing mode vibrations are themselves out of plane and vary significantly as a function of layer number. The relative atomistic motions of the SMs and LBMs in TMDs are illustrated in Figure 1, whereby the SM involves the in-plane motion of metal and chalcogen atoms, and the LBM involves the out-of-plane motion of metal and chalcogen atoms[43]. These SMs and LBMs are not present in single layers, but show a characteristic blue and red shift, respectively as layer number increases from 2L to bulk. While not commonly used as a metric for layer thickness in 2D materials currently, due their Raman shift position appearing in the ultra-low frequency region beyond the filter cut-offs for most commercial Raman spectrometers, ongoing developments in the use of components such as multiple notch filters[53] can allow measurement of these peaks with low excitation powers and short acquisition times. Full measurement and analysis of these modes is desirable for a more comprehensive understanding of the mechanical and electrical properties of TMDs[54].

The low-frequency SMs and LBMs have been extensively studied in graphene[53, 55] and have been reported for a number of mechanically exfoliated TMDs[30, 54, 56]. Unlike graphene, which consists of single atomic layers of carbon, TMD monolayers consist of three atomic layers of chalcogen/metal/chalcogen, resulting in a richer and more complex Raman spectrum. While previous reports have outlined the evolution of low-frequency peak positions with layer number[57], this has not yet been comprehensively studied for all layered materials. In addition, recent reports have identified new peaks in the low-frequency region of $MoSe_2$ corresponding to different polytypes of the material, indicating that Raman shifts in this region are of interest for considering differences in interlayer interactions with stacking type[33]. Here, by means of Raman mapping, we image the peak intensities and positions of SMs and LBMs for different TMDs and highlight the efficacy of this technique for layer-number identification. We further outline the effectiveness of this technique for quickly distinguishing between different stacking configurations which are prevalent in transition metal diselenide layers.

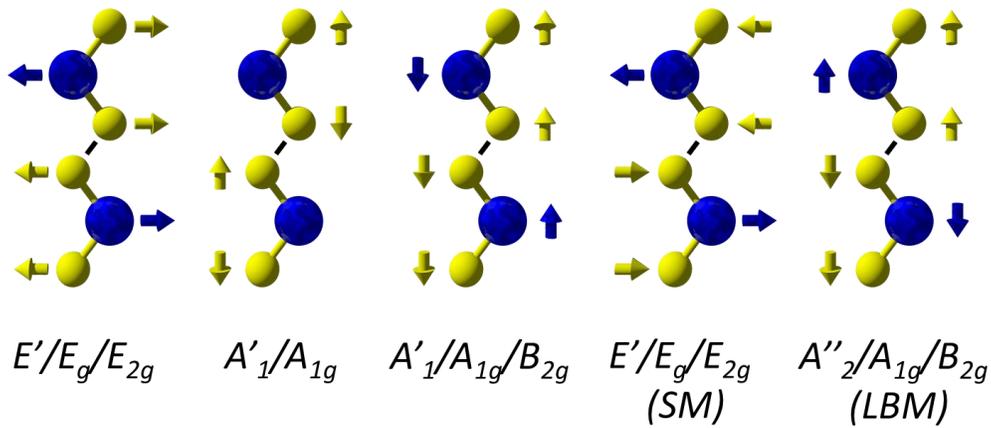

E'/$E_g$/$E_{2g}$   A'$_1$/$A_{1g}$   A'$_1$/$A_{1g}$/$B_{2g}$   E'/$E_g$/$E_{2g}$ (SM)   A''$_2$/$A_{1g}$/$B_{2g}$ (LBM)

Figure 1 – Schematic representation (ball and stick model) of Raman active modes in TMDs with the relative odd/even/bulk symmetry label indicated for each mode. Blue balls represent transition metal atoms; yellow balls represent chalcogen atoms, with arrows showing direction of motion.

## MoS$_2$ Raman Mapping

In Figure 2(a), a sample of CVD-grown MoS$_2$ with multiple distinct layers present is shown. In MoS$_2$, the in-plane (E'/$E_g$/$E^1_{2g}$) and out-of-plane (A'$_1$/$A_{1g}$) peaks occur in the vicinity of ~385 and ~403 cm$^{-1}$, respectively. Figure 2(d) shows the evolution of Raman spectra (normalized to A'$_1$/$A_{1g}$ peak intensity) extracted from 1-5L MoS$_2$, which display a characteristic red and blue shift of the E'/$E_g$/$E^1_{2g}$ and $A_1$'/$A_{1g}$ modes, respectively as the layer number increases[44, 46]. Peak intensity maps are presented in Figure 2(b) and (c), showing an increase in intensity of A'$_1$/$A_{1g}$ and $E'/E_g/E^1_{2g}$ peaks as layer number increases from 1-5 layers, with a subsequent decrease as layer number increases towards bulk, attributed to optical interference occurring for the excitation laser and emitted Raman scattering[46]. The peak position maps for $A'_1/A_{1g}$ and $E'/E_g/E^1_{2g}$ are shown in the supporting information, with Figure S1(a) and S1(b) showing clearly the red and blue shift in $E'/E_g/E^1_{2g}$ and $A'_1/A_{1g}$ peaks, respectively, as layer number increases, allowing an initial assessment of layer number to be made. This assessment is supported by PL intensity maps of the same area, shown in Figure 2(e) and (f), showing a maximum intensity of A1 excitons in monolayer regions, and an enhancement of B1 exciton intensity in multilayer regions. The corresponding shift in PL position as layer number increases, reflecting the changing bandgap of $MoS_2$ with layer number, is illustrated in the peak position maps in Figure S1(d) and (e) in the supplementary information, and in the corresponding spectra in Figure S1(f).

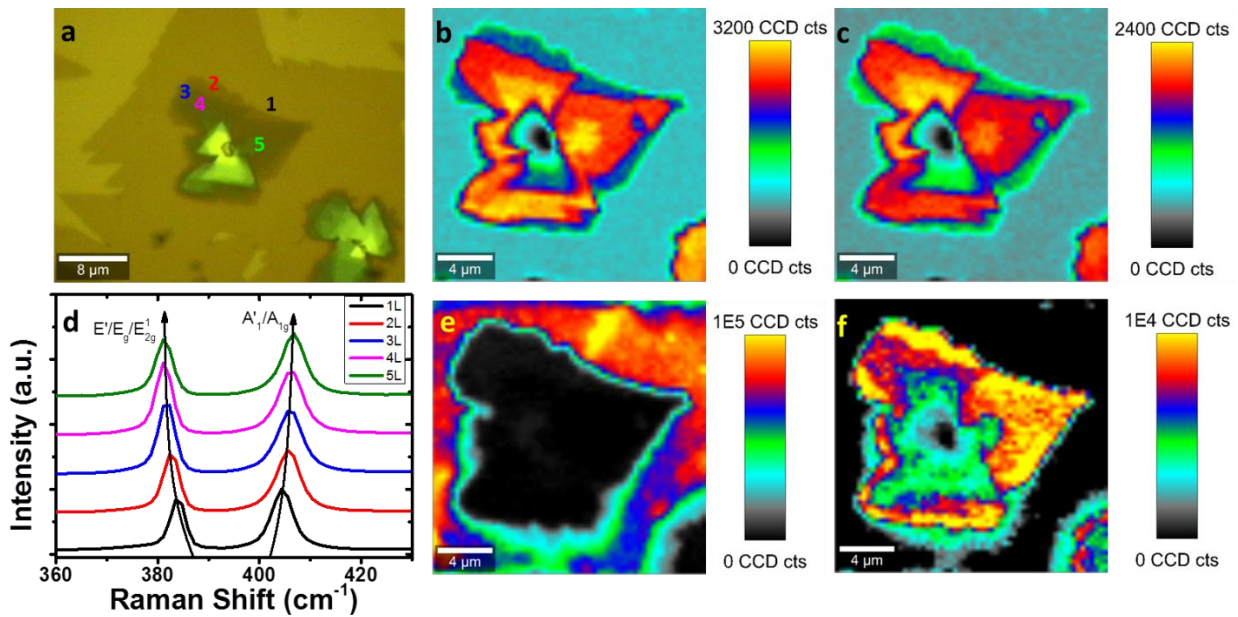

Figure 2 (a) Optical image of CVD $MoS_2$ with layer numbers labelled (b) Peak intensity map of $A'_1/A_{1g}$ (~403 cm$^{-1}$) high-frequency Raman mode (c) Peak intensity map of $E'/E_g/E^1_{2g}$ (~385 cm$^{-1}$) Raman mode (d) Raman spectra of 1-5L $MoS_2$ (e) Peak intensity map of A1 exciton PL peak (f) Peak intensity map of B1 exciton PL peak.

Figure 3 presents the low-frequency SMs and LBMs of $MoS_2$. Spectra of 1-5L $MoS_2$ are shown in Figure 3(a), in close agreement with previous measurements of mechanically exfoliated $MoS_2$[30, 54]. Figures 3(b to e) show peak intensity maps of SMs/LBMs for 2-5L $MoS_2$. There is some overlap in peak intensity maps, due to peaks for different layer numbers appearing at similar Raman shifts; however, the relative intensity of these modes provides a strong indication of layer number. While peak intensity maps allow a step-by-step assignation of layer number, this can be better visualized by generating a map of the position of maximum peak intensity in the low-frequency regime as shown in Figure 3(f). Such mapping represents a clear and facile method of assigning the layer number present in $MoS_2$, by uniquely identifying the highest intensity SMs and LBMs present in 2-5L $MoS_2$ by their position in the range of 10-50 $cm^{-1}$, noting that 1L $MoS_2$ has no peaks in this region.

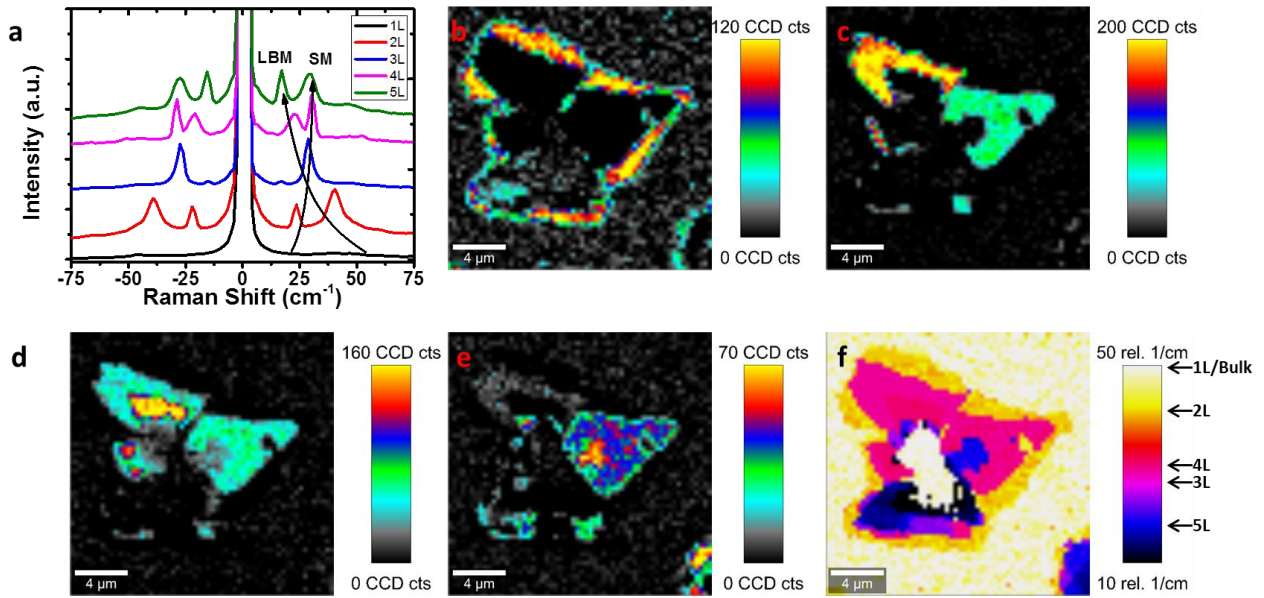

Figure 3 - (a) Low-frequency Raman spectra of SMs and LBMs of 1, 2, 3, 4 and 5L $MoS_2$ (b) Peak intensity map of LBM mode for 2L $MoS_2$ at ~40 $cm^{-1}$ (c) Peak intensity map of max SM/LBM for 3L $MoS_2$ at ~29 $cm^{-1}$ (d) Peak intensity map of max SM for 4L $MoS_2$ at ~31 $cm^{-1}$ (e) Peak intensity map of LBM for 5L $MoS_2$ at ~17 $cm^{-1}$ (f) Map of position of maximum peak intensity in the region of 10-50 $cm^{-1}$.

## MoSe$_2$ Raman Mapping

Raman analysis of CVD-grown MoSe$_2$ with a variety of layer numbers is shown in Figure 4. In MoSe$_2$, the in-plane ($E'/E_g/E^1_{2g}$) and out-of-plane ($A'_1/A_{1g}$) Raman active modes occur in the vicinity of ~287 and ~240 cm$^{-1}$, respectively. The significant red shift of peaks compared with MoS$_2$ occurs due to the larger mass of the selenium vs. sulfur atoms[54]. Similar to MoS$_2$, the in-plane ($E'/E_g/E^1_{2g}$) and out-of-plane ($A'_1/A_{1g}$) modes exhibit a red and blue shift, respectively, with increasing layer thickness. In Figure 4(a), an optical image of CVD grown layers is shown. A Raman map of $A'_1/A_{1g}$ (~240 cm$^{-1}$) peak intensity is shown in Figure 4(b), with the corresponding peak position map in Figure S2(a) in the supporting information. It is clear from these images that while the intensity varies significantly with thickness, following an initial jump from 1 to 2L, the $A'_1/A_{1g}$ (~240 cm$^{-1}$) position does not change dramatically with layer number. A map of the $E'/E_g/E^1_{2g}$ (~287 cm$^{-1}$) intensity is shown in Figure 4(c), with the corresponding position map in Figure S2(b) in the supporting information. This Raman mode's intensity and position changes significantly from monolayer to bilayer, but shows no further significant change between 2, 3, and 4 layers, and is therefore not useful for layer number determination. Figure 4(d) shows spectra of 1 to 4L 2H MoSe$_2$ crystals extracted from different areas in Figure 4(a), which are in good agreement with previously reported spectra[42, 47, 56]. We can also consider the intensity maximum and position maps of the $A'_1/A_{1g}/B^1_{2g}$ mode (~350 cm$^{-1}$). This mode is inactive in bulk material, but has previously been observed to become weakly Raman active in bilayer and few-layer crystals due to the breakdown of translation symmetry[42]. To avoid confusion with other modes, this will henceforth be referred to as the $B^1_{2g}$ mode. As this mode does not appear for monolayer MoSe$_2$, as shown in the peak intensity map in Figure 4(e), its absence (in combination with a characteristic PL signal) serves as a confirmation of monolayer presence. However, similar to $E'/E_g/E^1_{2g}$ (~287 cm$^{-1}$), it does not shift significantly in intensity or position for 2+ layers as shown in the map of $B^1_{2g}$ position in Figure S2(d) of the Supporting Information. A map of PL intensity is shown in Figure 4(f), with the corresponding position map and spectra shown in Figure S2(e) and (f), respectively in the Supporting Information. The intense PL seen in certain areas serves as confirmation of monolayer presence, with some drop-off in intensity, as expected, in the regions of grain boundaries. The apparent lack of PL in other layers does not necessarily signify bulk behaviour – rather the signal for few-layer crystals is overshadowed by that of the monolayer.

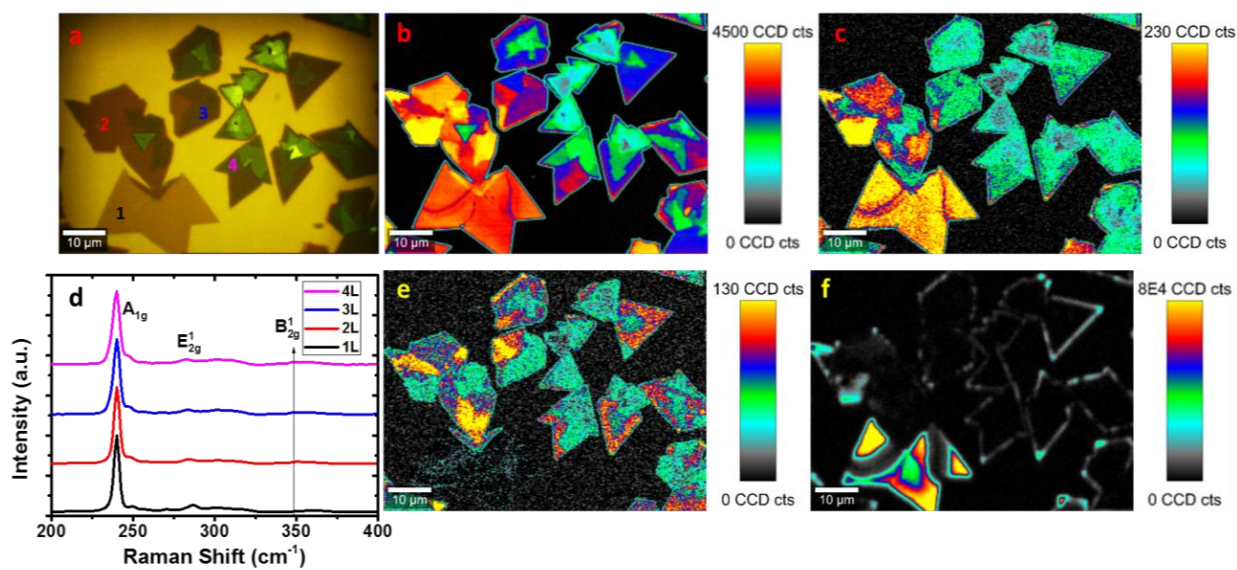

Figure 4 - (a) Optical image of CVD-grown MoSe$_2$ with varying layer numbers (b) Peak intensity map of $A'_1/A_{1g}$ (~ 240 cm$^{-1}$) Raman mode for MoSe$_2$ (c) Peak intensity map of $E'/E_g/E^1_{2g}$ (~287 cm$^{-1}$) Raman mode (d) Raman spectra of 1-5 L MoS$_2$ normalized to $A'_1/A_{1g}$ mode intensity (e) Peak intensity map of $B^1_{2g}$ (~350 cm$^{-1}$) mode (f) PL intensity map.

We now focus on the study of low-frequency Raman modes in MoSe$_2$. Figure 5(a) shows spectra of 1 to 4L 2H-MoSe$_2$ which have been extracted from different areas marked in the optical image in Figure 4(a), and are in close agreement with spectra previously shown in the literature[47]. These spectra have been normalized to the intensity of the high-frequency $A_{1g}$ mode, and offset for clarity, as have the rest of the MoSe$_2$ spectra in Figure 5. A Raman map of the maximum signal over the range 10-50 cm$^{-1}$ is shown in Figure 5(b). Interestingly, this map shows fractures and splitting in areas where no change is discernible in the optical image and therefore further investigation into the low-frequency modes was warranted. By analysis of various regions that appeared to be the same thickness according to optical contrast, it was possible to extract different low-frequency Raman signals correlating to different combinations of 2H and 3R stacking of MoSe$_2$ layers. These different stacking configurations have previously been observed in CVD-grown transition metal diselenide layers and their formation attributed to the small difference in formation energy between the two different configurations[58]. It should be noted that there was no evidence of these different stacking configurations in our CVD-grown MoS$_2$, with all areas probed displaying a purely 2H signal. In Figure 5(c), a map of position of peak intensity maximum in the low-frequency region is shown. Study of the differences in intensity maximum in Figure 5(b) and the position of this intensity maximum for each layer shows that there is no direct overlap in each – rather, some areas have peaks of maximum intensity in the same position but of different intensity, while others have peaks of similar intensity but in different areas. To explain this observation, we will examine the low-frequency spectra for each layer number. In Figure 5(d), low-frequency Raman spectra for different regions of 2L MoS$_2$ are shown, corresponding to 2H (max at 18 cm$^{-1}$), 3R (max at 18 cm$^{-1}$, but significantly lower in relative intensity), and 3R* (max at 29 cm$^{-1}$). The difference between 3R (max at 18 cm$^{-1}$) and 3R* (max at 29 cm$^{-1}$) is attributed to one being 3R, and the other being the vertically flipped 3R[33], labelled as 3R* here, which would interact radically differently with incoming phonons. The intensity maximum for 2H and 3R (18 cm$^{-1}$) is shown in Figure 5(e), which shows (with some overlap with peaks present in 4L) the areas where these peaks are present. The difference in intensity between 2H and 3R here is consistent with previous reports[33]. Additionally, as shown in Figure 5(f), we also observe experimentally for the first time a predicted Raman mode at ~29 cm$^{-1}$, attributed to the $A_1$ mode in the 3R* stacking configuration[33]. Similar evidence for different stacking configurations is seen in the 3L low-frequency Raman spectra in Figure 5(g), where it is possible to identify a variety of 3L stacking configurations, including 2H-2H, 2H-3R, and 3R-3R. The trends in intensity for the peaks at ~13 cm$^{-1}$ and ~24 cm$^{-1}$ are clear when the peak intensity maps are considered. In Figure 5(h), a peak intensity map of the SM at ~13 cm$^{-1}$ is shown, which is present for 3R-3R stacking, but also present at higher intensities as the SM mode in 2H-3R stacking, where it appears in parallel with another SM mode at 24 cm$^{-1}$. Therefore, the relative intensity of this mode at ~13 cm$^{-1}$ can be used to distinguish between 3R-3R and 2H-3R stacking, as labelled on the intensity scale bar in Figure 5(h), with further verification of the 2H-3R mode afforded by the presence of a SM/LBM overlap peak at ~24 cm$^{-1}$, the intensity of which is mapped out in Figure 5(i). This peak is highest in intensity in 2H-2H stacking, as is expected for pristine mechanically exfoliated 2H crystals[56], and decreases as stacking configuration goes from 2H-2H to 2H-3R to 3R-3R. This is logical when considering the decreasing interlayer interactions and force constants present in 3R stacking in comparison to 2H stacking. The respective intensities for the different stacks, as shown in Figure 5(i), indicate clearly that different intensities are present for this peak in different areas, allowing one to distinguish between 2H-2H, 2H-3R and different 3R-3R stacking configurations. The use of Raman intensity maps serves to highlight the ubiquitous nature of the different stacking configurations, which would not be readily apparent in comparing individual spectra of different crystals, or in the study of high-frequency point spectra, which show little change between 2H and 3R stacking configurations[32], as shown in the extracted high-frequency spectra in Figure S4 and discussed in the Supporting Information. Low-frequency Raman mapping can distinguish between different stacking configurations rapidly and non-destructively, allowing TMDs in different stacking configurations to be identified and studied without the need for high-resolution imaging[59]. The peak positions of SMs and LBMs observed here are in good agreement with previously observed low-frequency modes in mechanically exfoliated 2H $MoSe_2$[56, 60] and CVD-grown $MoSe_2$ stacking polytypes[33]. Raman spectra of different stacking configurations for 4L $MoSe_2$ are shown in Figure S3 and discussed in the Supporting Information. Layer number assignations have been confirmed using atomic force microscopy (AFM) as detailed in the Supporting Information, Figures S5 and S6.

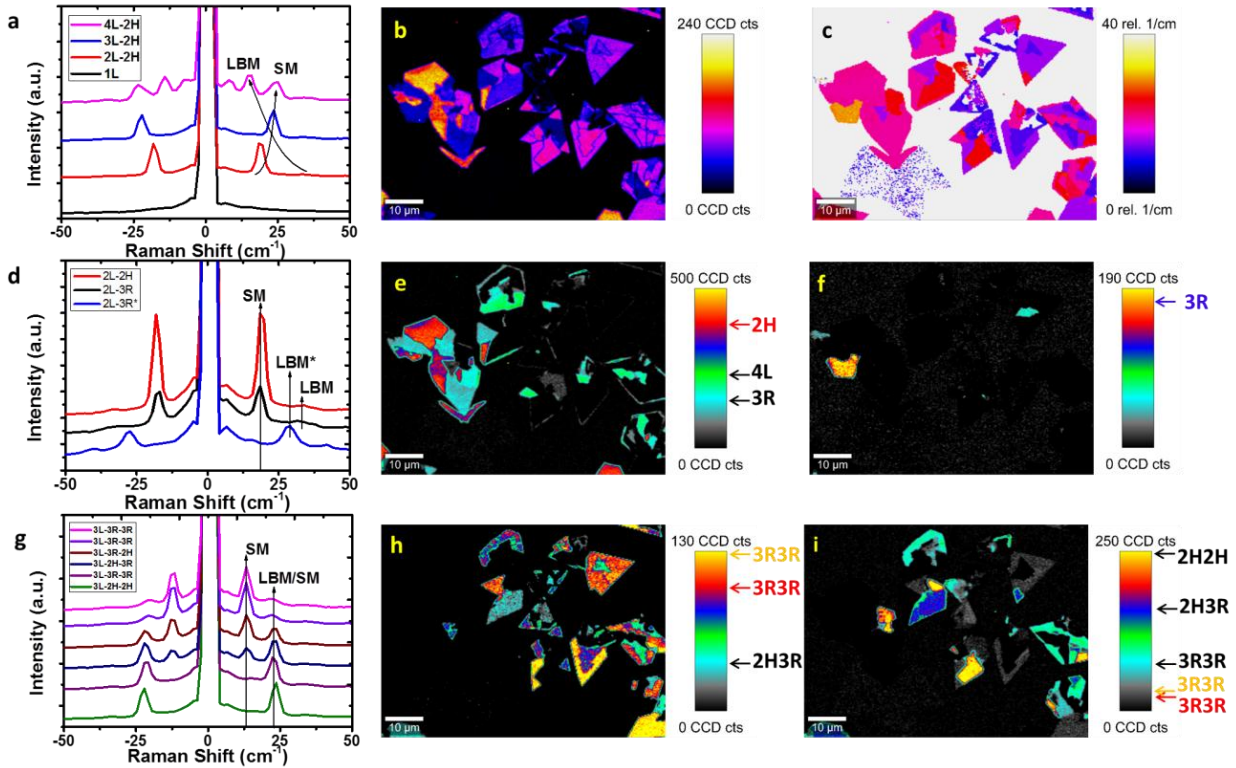

Figure 5 - (a) Low-frequency Raman spectra of SMs and LBMs of 1, 2, 3, and 4L 2H $MoSe_2$ (b) Peak intensity map over the range 10-50 $cm^{-1}$ (c) Map of position of maximum peak intensity of the low-frequency Raman modes in the range of 0-40 $cm^{-1}$ (d) Low-frequency Raman spectra of SMs and LBMs of 2H and 3R stacking configurations in 2L $MoSe_2$ (e) Peak intensity map for 2L $MoSe_2$ SM at ~18 $cm^{-1}$ (f) Peak intensity map for 2L $MoSe_2$ LBM at ~29 $cm^{-1}$ (g) Enhanced low-frequency Raman spectra of SMs and LBMs of 2H and 3R combination stacking configurations for 3L $MoSe_2$ (h) Peak intensity map for 3L $MoSe_2$ at ~13 $cm^{-1}$ (i) Peak intensity map for 3L $MoSe_2$ at ~24 $cm^{-1}$

**WSe₂ Raman Mapping**

A sample of CVD-grown WSe$_2$ with a variety of layer numbers present is shown in Figure 6(a). The WSe$_2$ Raman spectrum displays the in-plane ($E'/E_g$) and out-of-plane ($A'_1/A_{1g}$) modes typical for layered TMDs. Under the experimental conditions used here, these appear as a single overlapping peak at ~250 cm$^{-1}$ in mono- and few-layer WSe$_2$. In the case of resonant excitation conditions, as applies when using a 532 nm excitation laser in resonance with the A' exciton peak of WSe$_2$[61, 62], the *2LA(M)* phonon also appears. This is a second order resonant Raman mode that occurs due to *LA* phonons at the M point in the Brillouin zone[45], similar to the case of MoS$_2$ and WS$_2$ in resonance[26, 63, 64]. Figure 6(d) shows spectra of 1 to 3L WSe$_2$ extracted from different areas marked in Figure 6(a), which are in agreement with previous studies[56, 58]. A peak intensity Raman map of the peak at ~250 cm$^{-1}$ is shown in Figure 6(b), with the corresponding position map in Figure S7(a) in the Supporting Information. This peak is a combination of contributions from the $A'_1/A_{1g}$ and $E'/E_g/E^1_{2g}$ modes that coincidentally overlap at this Raman shift. This mode shows a decrease in intensity with layer number, and a slight shift in position as shown and discussed in Figure S7(a) in the Supporting Information. The changing intensity of this peak between the two bilayer regions, as labelled on the optical image, indicates some change in stacking configuration, with one region appearing at a higher intensity than the other[58]. This is likely due to a decrease in in-plane contributions due to decreasing magnitudes of Raman tensors in 3R symmetry contributions, but high-frequency modes alone are not sufficient to assign a definitive stacking configuration to each region. The labels shown on the optical image will be discussed in the low-frequency analysis below. A Raman map of the *2LA(M)* mode (~260 cm$^{-1}$) intensity is shown in Figure 6(c), with the corresponding position map in Figure

S7(b). The *2LA(M)* mode's intensity changes significantly from monolayer to bilayer, but shows no further significant change for 3L. It is clear that this mode, similar to the peak at 250 cm$^{-1}$, is also more intense for one bilayer region than another. The relative intensity of *2LA(M)* increases with respect to the *A'$_1$/A$_{1g}$* and *E'/E$_g$/E$^1_{2g}$* combination peak, however, the overall intensity decreases sufficiently for this not to be apparent in the peak intensity maps. The *B$_{2g}$* (~310 cm$^{-1}$) peak intensity map is shown in Figure 6(e), with the corresponding peak position map shown in Figure S7(d). This mode, similar to the case for MoSe$_2$, is inactive in bulk material, but becomes Raman active in few-layer samples[42]. However, the absence of a discernible change in the intensity or position for 2-3 layers means it is of little use for layer-number analysis. Interestingly, this mode is most intense in the case of one 2L stacking configuration, which we tentatively attribute to increased interlayer interactions in ideal (likely 2H stacking) in comparison to other (3R) configurations. The brightest areas in the PL intensity map in Figure 6(f) signify the presence of monolayers. This is confirmed by the extracted PL spectra and position map shown in Figure S7(c) and (e), respectively in the Supporting Information. As layer number increases, the PL position shifts to higher wavelengths (lower bandgap), and decreases in intensity, as is expected due to the change in band structure[1, 63]. No significant change in PL intensity or position is seen between the two different bilayer regions.

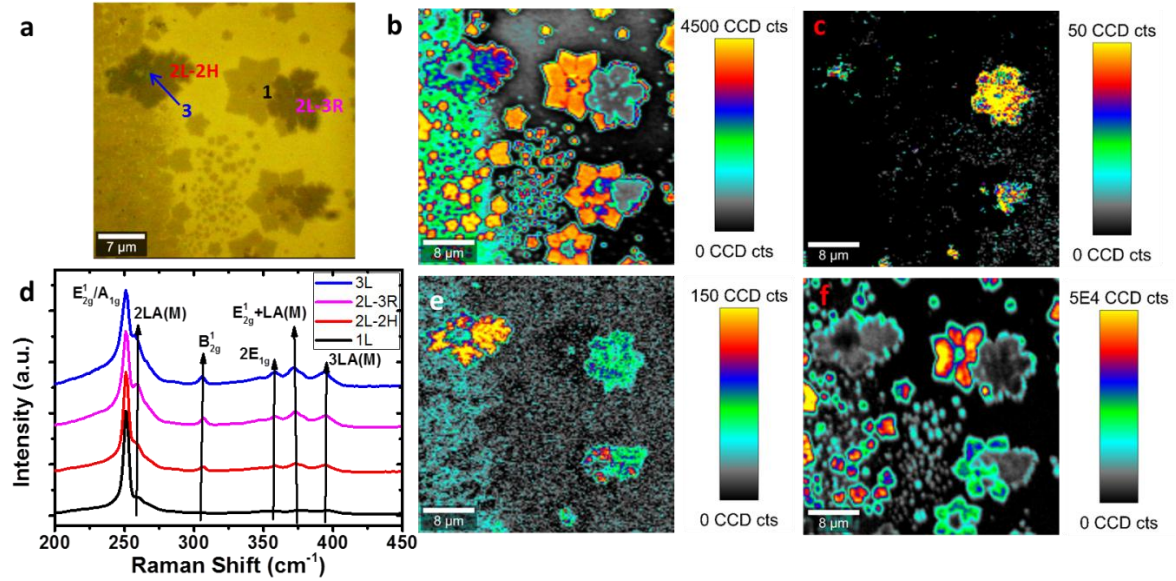

Figure 6 - (a) Optical image of CVD WSe$_2$ with varying layer numbers (b) Peak intensity map of $A'_1/A_{1g}$ and $E'/E_g/E^1_{2g}$ overlapping modes (~250 cm$^{-1}$) (c) Peak intensity map of *2LA(M)* peak (~260 cm$^{-1}$) (d) Raman spectra of 1L, 2L-2H, 2L-3R and 3L-2H WSe$_2$ (e) Peak intensity map of $A'_1/A_{1g}/B_{2g}$ (~310 cm$^{-1}$) Raman mode (f) PL intensity map.

The low-frequency Raman modes of WSe$_2$ are shown in Figure 7. Figure 7(a) shows spectra of 1 to 3L WSe$_2$ SMs and LBMs, which have been extracted from different areas marked in the optical image in Figure 6(a), and are in close agreement with spectra previously reported[56, 58]. A clear decrease in intensity of the SM from 2L-2H to 2L-3R stacking is observed, with a corresponding increase in the LBM. The low-frequency peaks shown here agree well with different stacking configurations of 2L WSe$_2$ reported previously[58]. A Raman map of the 2L SM (~17 cm$^{-1}$) intensity is shown in Figure 7(b), which shows (with some overlap with peaks present in different layers) the areas where 2L-2H coverage is present. This is also shown for intensity maps of 2L-3R LBM (~27 cm$^{-1}$) and 3L-2H SM/LBM peak overlap (~21 cm$^{-1}$) shown in Figure 7(c) and (d), respectively. A map of the position of maximum intensity in the low-frequency region is shown in Figure 7(e), where the measurement of peak position over the range of 10-40 cm$^{-1}$ allows for some clarification of each layer from a single Raman map.

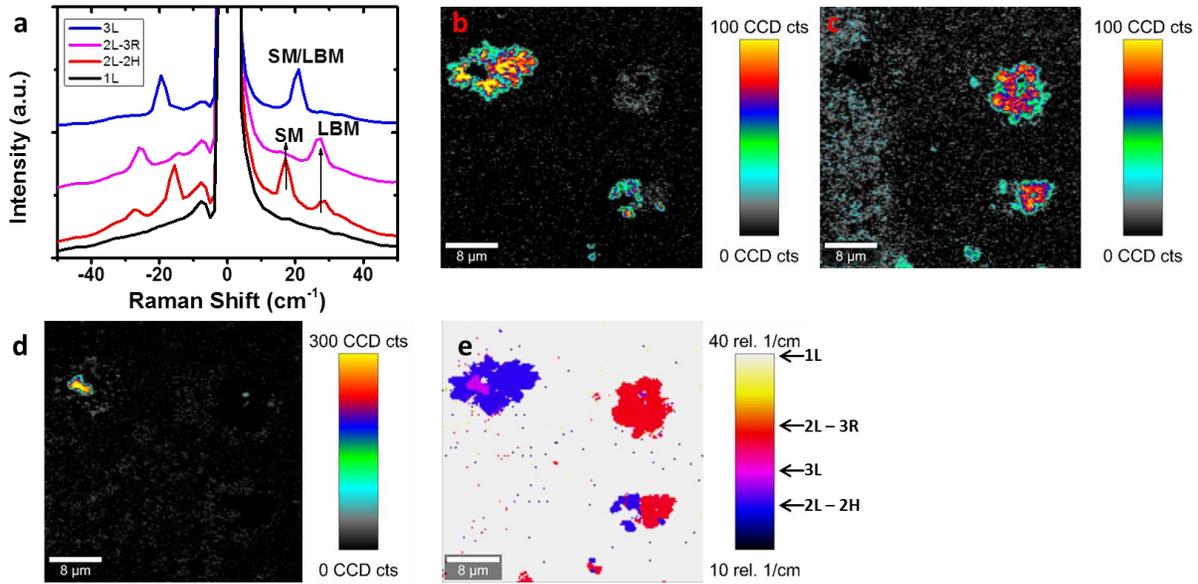

Figure 7 - (a) Low-frequency Raman spectra of SMs and LBMs of 1, 2 and 3L WSe$_2$ (b) Peak intensity map of SM mode for 2L-2H WSe$_2$ at ~17 cm$^{-1}$ (c) Peak intensity map for 2L-3R WSe$_2$ at ~27 cm$^{-1}$ (d) Peak intensity map of SM/LBM mode for 3L-2H WSe$_2$ at ~21 cm$^{-1}$ (e) Map of position of maximum peak intensity of the low-frequency Raman modes in the range of 10-40 cm$^{-1}$.

## WS$_2$ Raman Mapping

A sample of CVD grown WS$_2$ with a variety of layer numbers present is shown in Figure 8(a). The WS$_2$ Raman spectrum with an excitation wavelength of 532 nm is characterized by the $E'/E_g/E^1_{2g}$ and $A'_1/A_{1g}$ modes at ~355 cm$^{-1}$ and 417 cm$^{-1}$, respectively, and the resonant *2LA(M)* phonon mode at ~352 cm$^{-1}$, similar to that discussed previously for WSe$_2$. The resonance mode appears here due to the 532 nm laser wavelength used being in resonance with the B exciton peak of WS$_2$[61, 62, 64]. Resonant Raman spectroscopy is a powerful tool in the study of exciton-phonon interactions in 2D materials; through careful selection of the excitation wavelength certain modes can be enhanced and additional resonant contributions such as the *2LA(M)* mode observed[65]. A Raman map of intensity of the peak centred at ~352 cm$^{-1}$ is shown in Figure 8(b), with the corresponding peak position map in Figure S8(a) in the supporting information. This peak is a combination of contributions from the resonant *2LA(M)* and $E'/E_g/E^1_{2g}$ modes that coincidentally overlap at this Raman shift. This peak is most intense in monolayer crystals, correlating to the PL map in Figure 8(e). A Raman map of the $A'_1/A_{1g}$ mode intensity is shown in Figure 8(d), with the corresponding peak position map shown in Figure S8(b) in the Supporting Information. The Raman spectrum of these layers is shown in Figure 8(c), with the spectra normalized to the peak at 352 cm$^{-1}$ and offset for clarity. This shows changing behaviour from monolayer to few-layer crystals that is consistent with previous reports[26, 66]. The remarkable PL in WS$_2$ monolayers is evident in the PL intensity map and spectrum in Figure S8(e) and (g) in the Supporting Information. The apparent absence of PL in this map for 2+ layers is simply due to the relative intensity of the PL in 2+ layers being dwarfed by the emission from the monolayer crystals, where the intensity ratio of PL to $2LA(M)/E^1_{2g}$ is ~25. Further changes in PL between mono and few-layer films are evident in the map of PL peak position in Figure S8(f) in the Supporting Information, which demonstrates the position shift from ~640 nm for monolayers to ~650 nm for few layers, as is expected as the addition of layers causes shifting of the band structure towards a smaller and more indirect bandgap.

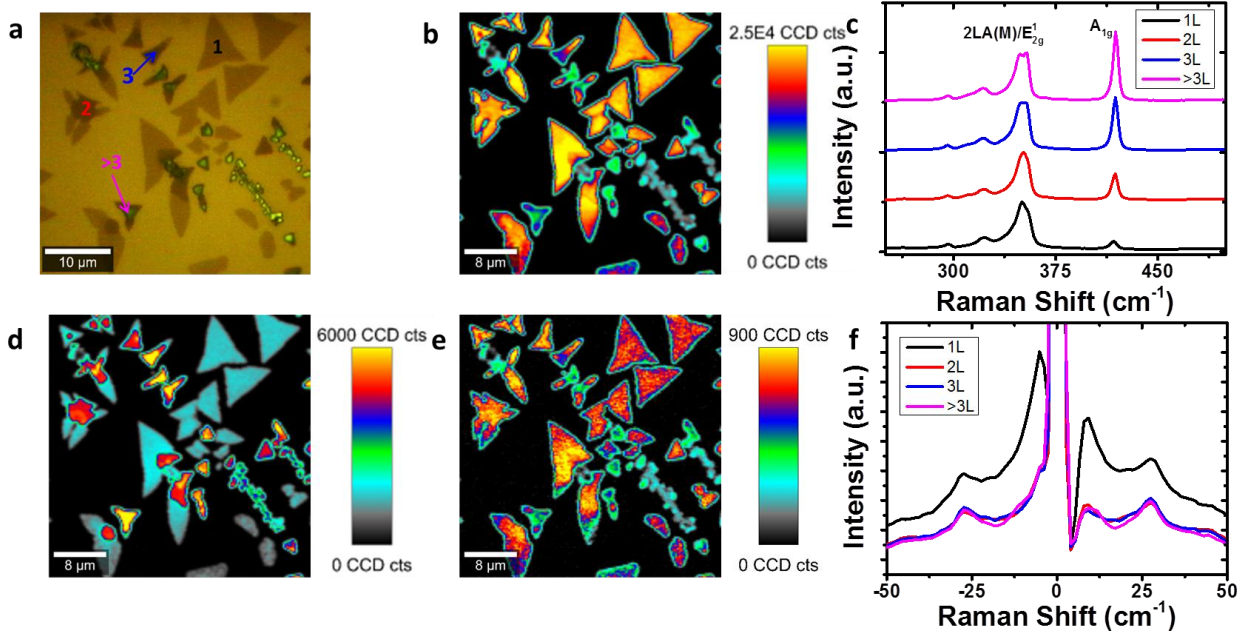

Figure 8 - (a) Optical image of CVD WS$_2$ with varying layer numbers (b) Peak intensity map of $2LA(M)+E'/E_g/E^1_{2g}$ (~352 cm$^{-1}$) (c) Raman spectra of 1, 2, 3, and >3L WS$_2$ in the high-frequency region (d) Peak intensity map of max

*A'₁/A₁g* peak (~417 cm⁻¹) (e) Peak intensity map of low-frequency resonance mode at 27 cm⁻¹ in WS$_2$ (f) Raman spectra of 1, 2, 3, and 4L WS$_2$ in the low-frequency region

While for MoS$_2$, MoSe$_2$ and WSe$_2$ we have highlighted the practicality of low-frequency Raman spectroscopy for assessment of layer-number and stacking orientation, in the case of WS$_2$ we will now discuss the possible presence of resonant modes in the low-frequency region of the Raman spectrum. Low-frequency Raman spectra of WS$_2$ regions of different layer thickness are shown in Figure 8(f). We observe a peak at ~27 cm⁻¹ for all layer numbers, essentially obscuring SMs and LBMs at the Raman excitation wavelength used (532 nm). This peak is most intense in monolayer, as can be seen by the map in Figure 8(e). A recent report has shown similar behaviour in the low-frequency region of the Raman spectrum of MoS$_2$ probed with a 633 nm excitation laser[67] and attributed this to strong resonance with excitons or exciton-polaritons, while previous reports have attributed this resonant Raman process to be reflective of a subtle splitting in the conduction band at *K* points[68]. We tentatively assign this new peak in WS$_2$ as a *LA(M)* related mode, due to the peak intensity maps appearing almost identical in relative intensity to the *2LA(M)* peak intensity map shown in Figure 8(b). It should be noted that these resonance effects are not seen in WSe$_2$, with the laser wavelength used (532 nm), as this is only in resonance with the A' split exciton peak, and not an exciton absorption peak as is the case for WS$_2$[61]. This peak is seen in WS$_2$ for all layer thicknesses measured and while it is most intense in monolayer it does not vary significantly in intensity for other layer numbers. To further strengthen the link between this newly observed peak and the resonant modes, a comparison between Figure S8(c) in the Supporting Information, a peak intensity map of the *LA(M)* mode, and the low-frequency resonance peak shown here in Figure 8(e), shows that these correlate in relative intensity. It is suggested that further exploration of WS$_2$ low-frequency modes with multiple wavelengths would confirm this assignment, as has held true for MoS$_2$[67, 68].

## Conclusion

A comprehensive study of Raman scattering in CVD-grown mono- and few-layer $MoS_2$, $MoSe_2$, $WSe_2$ and $WS_2$ has been presented. Phonon modes for in-plane and out-of-plane vibrations show thickness dependent intensities and positions in both the high- and low-frequency regions. The general peak shift trends are similar for all materials studied due to their similar lattice structures, where a stiffening (blue shift) is observed in SMs, while a softening (red shift) is observed in LBMs, with increasing layer number. However, the intensity dependencies and Raman shifts vary in each material due to the different atomic masses of the metal/chalcogen in each crystal type, and due to the stacking order of the layers. The determination of layer number via systematic low-frequency mode mapping is a crucial development in the research and analysis of TMD thin films, as is the stacking configuration determination, which we have shown here by Raman mapping techniques. We further report a new peak observable in resonance conditions at ~27 cm$^{-1}$ in $WS_2$ crystals.

In future, low-frequency Raman mapping could readily be applied to quickly assess the layer number of TMDs produced by other methods, such as liquid-phase exfoliation, to ascertain their suitability for specific applications. Importantly, this methodology could be extended to other TMD crystals that do not show significant changes in the high-frequency region of their Raman spectrum with layer number, such as $ReS_2$[69]. Furthermore, it is anticipated that this technique will be useful for investigating layer number and stacking orientation in 2D material alloys[70] and recently fabricated TMD heterostructures[39, 40].

## Materials and Methods

### CVD growth of TMDs

Precursor layers of $MoO_3$ ($WO_3$) were liquid-phase exfoliated and dispersed onto commercially available silicon dioxide ($SiO_2$, ~290 nm thick) substrates as described previously[28, 71]. The $MoO_3$ ($WO_3$) precursor substrates were then placed in a quartz boat with a blank 300 nm $SiO_2$/Si substrate face down on top of them, creating a microreactor. This was then placed in the centre of the heating zone of a quartz tube furnace, and ramped to 750 °C under 150 sccm of forming gas (10% $H_2$ in Ar) flow at a pressure of ~0.7 torr. Sulfide and selenide films were grown in separate, dedicated systems to avoid cross contamination.

For $MSe_2$ growth, Se vapour was then produced by heating Se powder to ~220 °C in an independently controlled upstream heating zone of the furnace, and carried downstream to the microreactor for a duration of 30 minutes after which the furnace was cooled down to room temperature.

For $MS_2$ growth, S vapour was then produced by heating S powder to ~120 °C in an independently controlled upstream heating zone of the furnace, and carried downstream to the microreactor for a duration of 20 minutes after which the furnace was held at 750 °C for 20 minutes before being cooled down to room temperature.

A schematic of the growth setup used is shown in Figure S9 in the Supporting Information. While the described growth procedure can produce large-area monolayer coverage[28], areas consisting of crystals with a variety of layer thicknesses were specifically chosen to highlight the capability of low-frequency Raman mapping for layer-number and stacking-orientation investigation.

**Raman and PL Analysis**

Raman and PL spectroscopy were performed using a Witec alpha 300R with a 532 nm excitation laser and a laser power of < 500 μW, in order to minimize sample damage. The Witec alpha 300R was fitted with a Rayshield Coupler to detect Raman lines close to the Rayleigh line at 0 cm$^{-1}$. A spectral grating with 1800 lines/mm was used for all Raman spectra whereas a spectral grating with 600 lines/mm was used for PL measurements. The spectrometer was calibrated to a Hg/Ar calibration lamp (Ocean Optics) prior to the acquisition of spectra. Maps were generated by taking 4 spectra per μm in both x and y directions over large areas. AFM measurements were carried out using a Veeco Dimension 3100 in tapping mode, with 40 N/m probes from Budget Sensors.

## Author Information

### Acknowledgements

This work is supported by the SFI under Contract No. 12/RC/2278 and PI_10/IN.1/I3030. M.O.B. acknowledges an Irish Research Council scholarship via the Enterprise Partnership Scheme, Project 201517, Award 12508. N.M. acknowledges SFI (14/TIDA/2329). The authors thank Christian Wirtz for illustrations as well as Riley Gatensby and Kangho Lee for assistance with CVD.

### Contributions

N.M. conceived and designed the experiments. N.M. and M.O. synthesized materials by CVD, carried out spectroscopic measurements and analysis and wrote the paper. D.H. and J.N.C. carried out liquid-phase exfoliation of precursor nanosheets. T.H. performed AFM measurements. N.M. and G.S.D. supervised the whole project. All authors contributed to the discussion of the results and improvement of the manuscript. †These authors contributed equally.

**Competing financial interests**

The authors declare no competing financial interests.

**Corresponding Author**

Correspondence to: Niall McEvoy (nmcevoy@tcd.ie), Georg S. Duesberg (duesberg@tcd.ie)

## References

1. Mak, K.F., Lee, C., Hone, J., Shan, J. & Heinz, T.F. Atomically Thin $MoS_2$ :A New Direct-Gap Semiconductor. *Phys Rev Lett* **105**, 136805 (2010).
2. Splendiani, A. et al. Emerging photoluminescence in monolayer $MoS_2$. *Nano Lett* **10**, 1271-1275 (2010).
3. Radisavljevic, B., Radenovic, A., Brivio, J., Giacometti, V. & Kis, A. Single-layer $MoS_2$ transistors. *Nat. Nanotechnol.* **6**, 147-150 (2011).
4. Lembke, D. & Kis, A. Breakdown of High-Performance Monolayer $MoS_2$ Transistors. *ACS nano* **6**, 10070-10075 (2012).
5. Lopez-Sanchez, O. et al. Light Generation and Harvesting in a van der Waals Heterostructure. *ACS Nano* **8**, 3042-3048 (2014).
6. Lopez-Sanchez, O., Lembke, D., Kayci, M., Radenovic, A. & Kis, A. Ultrasensitive photodetectors based on monolayer $MoS_2$. *Nat. Nanotechnol.* **8**, 497-501 (2013).
7. Yim, C. et al. Heterojunction Hybrid Devices from Vapor Phase Grown MoS2. *Scientific reports* **4**, 5458 (2014).
8. Nolan, H. et al. Molybdenum disulfide/pyrolytic carbon hybrid electrodes for scalable hydrogen evolution. *Nanoscale* **6**, 8185-8191 (2014).
9. Acerce, M., Voiry, D. & Chhowalla, M. Metallic 1T phase $MoS_2$ nanosheets as supercapacitor electrode materials. *Nat Nano* **10**, 313-318 (2015).
10. Wang, T. et al. Size-Dependent Enhancement of Electrocatalytic Oxygen-Reduction and Hydrogen-Evolution Performance of $MoS_2$ Particles. *Chemistry – A European Journal* **19**, 11939-11948 (2013).
11. Eda, G. et al. Photoluminescence from Chemically Exfoliated $MoS_2$. *Nano Letters* **11**, 5111-5116 (2011).
12. Coleman, J.N. et al. Two-Dimensional Nanosheets Produced by Liquid Exfoliation of Layered Materials. *Science* **331**, 568-571 (2011).
13. Backes, C. et al. Edge and confinement effects allow in situ measurement of size and thickness of liquid-exfoliated nanosheets. *Nat Commun* **5**, 4576 (2014).

# Supplementary Information - Mapping of Shear and Layer-Breathing Raman modes in CVD-Grown Transition Metal Dichalcogenides: Layer Number, Stacking Orientation and Resonant Effects

Maria O'Brien[1,2†], Niall McEvoy[1,2†*], Damien Hanlon[2,3], Toby Hallam[2,3], Jonathan N. Coleman[2,3] and Georg S. Duesberg[1,2*]

[1]School of Chemistry, Trinity College Dublin, Dublin 2, Ireland

[2]Centre for Research on Adaptive Nanostructures and Nanodevices (CRANN) and Advanced Materials and BioEngineering Research (AMBER) Centre, Trinity College Dublin, Dublin 2, Ireland

[3]School of Physics, Trinity College Dublin, Dublin 2, Ireland

†These authors contributed equally.

**Additional MoS$_2$ Analysis**

Peak position maps for $A'_1/A_{1g}$ and $E'/E_g/E^1_{2g}$ modes are shown in Figure S1(a) and (b), respectively. A plot of the Raman shift position as a function of layer number in Figure S1(c) shows clearly the red and blue shift in $E'/E_g$ and $A'_1/A_{1g}$ peaks, respectively as layer number increases, allowing an initial assessment of layer number to be made, as indicated in the optical image in Figure 2(a) in the main text. The corresponding shift in PL position as layer number increases, reflecting the changing band structure of MoS$_2$ with layer number, is illustrated in Fig S1(d) and (e), which show the position of the A1 and B1 exciton, respectively. The extracted PL spectra in Figure S1(f) show the A1 and B1 excitonic peaks at ~680 and ~640 nm respectively, consistent with previous reports for CVD-MoS$_2$[1, 2]. The enhanced intensity in monolayers is due to their higher luminescence quantum efficiency than few-layer flakes. Inset in Figure S1(f) shows the PL normalized to the maximum signal for each layer number, to further emphasize the changing PL (reflecting the changing electronic properties) with layer number. This highlights the evolution in peak position and change in A:B exciton intensity ratio with changing layer thickness.

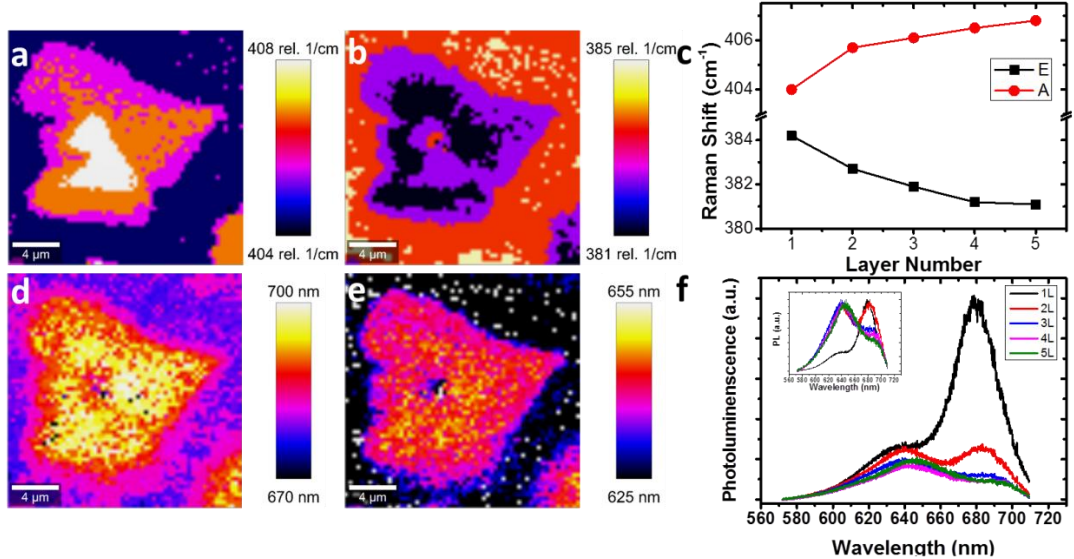

Figure S1 - Additional Raman and PL mapping of $MoS_2$ crystals of different layer thickness extracted from the regions shown in Figure 2(a) of the main text - (a) Map $A'_1/A_{1g}$ mode position (b) Map of $E'/E_g/E^1_{2g}$ mode position (c) Plot of position of Raman shift for $A'_1/A_{1g}$ and $E'/E_g/E^1_{2g}$ modes as a function of layer number. (d) Map of A1 exciton position (e) Map of B1 exciton position (f) PL spectra extracted from 1-5L regions corresponding to same areas as Raman spectra for each layer. Inset: Normalized version of the same spectra.

**Additional MoSe$_2$ Analysis**

Peak position maps for $A'_1/A_{1g}$ and $E'/E_g/E^1_{2g}$ modes are shown in Figure S2(a) and (b), respectively. In the peak position maps for $A'_1/A_{1g}$ in Figure S2(a), negligible variation is seen over the areas mapped, with the exception of expected decreases in the positions of grain boundaries[2], and variation at crystal edges, possibly due to doping or strain in these regions as has been reported for CVD-$MoS_2$[3, 4]. Figure S2(b) shows the peak position map for $E'/E_g/E^1_{2g}$, where a shift towards higher wavenumbers with decreasing layer numbers is observed. A slight blue-shift of $E'/E_g/E^1_{2g}$ is also seen at the edges of layers, likely due to edge and termination effects. Figure S2(c) shows a plot of Raman shift position for the $E'/E_g$ and $A'_1/A_{1g}$ modes as a function of layer number, with a significant red shift observed for $E'/E_g$ with increasing layer number, similar to MoS$_2$, but a less significant blue shift in $A'_1/A_{1g}$ with increasing layer number. Figure S2(d) shows the map of $B^1_{2g}$ mode position. This illustrates that this mode does not appear for monolayer, and while it is present for 2+L it does not vary significantly in position. The map of PL position for MoSe$_2$ is shown in Figure S2(e), showing a shift in position at the edge of each crystal, similar to previous observations for TMDs[3, 5]. The general shift in PL position for different regions is further shown in Figure S2(f), which shows the PL spectra extracted for the layer number regions discussed previously. The PL peak for monolayer occurs as single prominent maximum at a wavelength of ~800 nm, in line with previous reports of CVD and mechanically exfoliated MoSe$_2$[6, 7]. The shift in PL towards higher wavelengths (lower bandgaps) with increasing layer number is consistent with previous reports that show MoSe$_2$ and other 2D TMDs shifting to smaller and more indirect bandgaps with increasing layer numbers. Inset in Figure S2(f) shows the same PL spectra normalized to maximum PL intensity, in order to emphasize the shift in PL position with layer number.

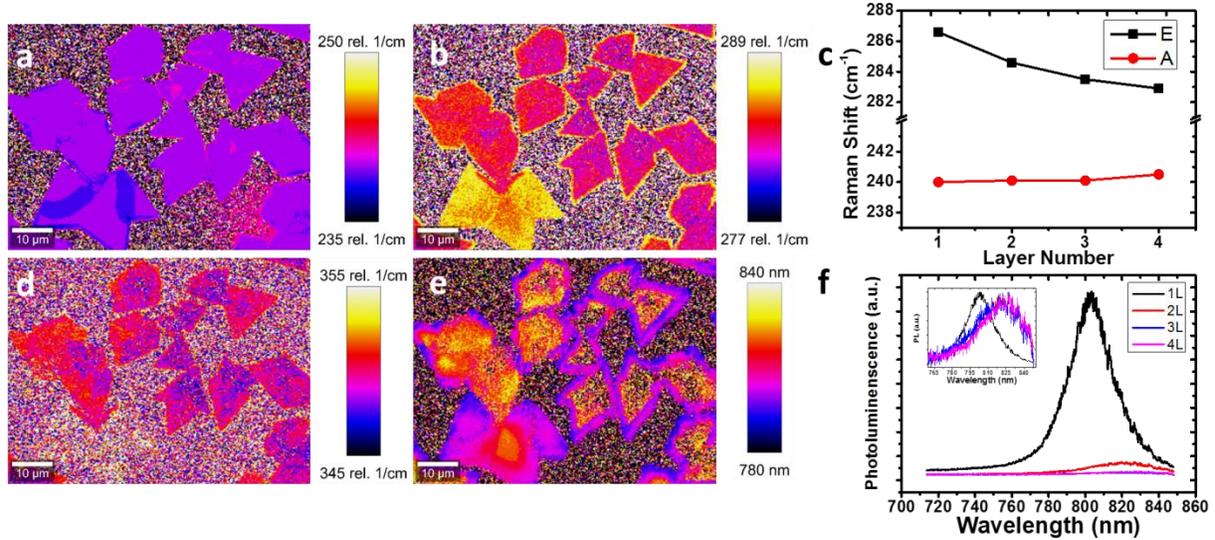

Figure S2 - Additional Raman and PL mapping of MoSe$_2$ crystals of different layer thickness extracted from the regions shown in Figure 4(a) of the main text - (a) Position map of maximum $A'_1/A_{1g}$ mode intensity (b) Position map of maximum $E'/E_g/E^1_{2g}$ mode intensity (c) Plot of position of Raman shift for $A'_1/A_{1g}$ and $E'/E_g/E^1_{2g}$ modes. (d) Position map of $B^1_{2g}$ mode intensity (e) PL position map (f) PL spectra of MoSe$_2$ crystals of different layer thickness extracted from the regions shown in Figure 4(a) of the main text. Inset: Normalized version of the same spectra highlighting the evolution in peak position with changing layer thickness.

Evidence for changing stacking configurations is seen in the 4L low-frequency Raman spectra in Figure S3, where it is possible to identify a variety of 4L stacking configurations, which we have labelled A-H. The characteristic modes at ~11 cm$^{-1}$, 18 cm$^{-1}$ and 24 cm$^{-1}$ are mapped in Figure S3(a), (b) and (c), with labels for each spectrum extracted on the intensity scale for each. To facilitate identification of the layers in this area, an optical image of Figure 3(a) with enhanced contrast applied is shown here in Figure S3(d). By analysis of the Raman maps in Figure S3(a)-(c), it was possible to extract spectra A-H shown in Figure S3(e).

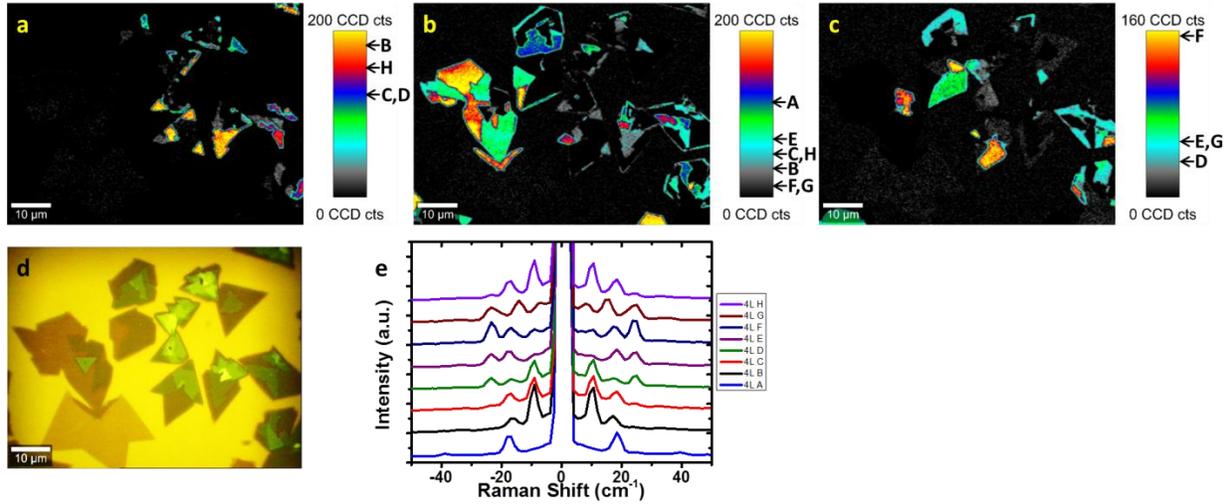

Figure S3 - Additional Raman mapping of MoSe$_2$ crystals of different layer thickness extracted from the regions shown in Figure 4(a) of the main text - (a) Peak intensity map for 4L MoSe$_2$ SM at ~11 cm$^{-1}$ (b) Peak intensity map for 4L MoSe$_2$ SM at ~18 cm$^{-1}$ (c) Peak intensity map for 4L MoSe$_2$ SM at ~24 cm$^{-1}$ (d) Optical image of the region of MoSe$_2$ crystals mapped with additional contrast applied (e) Spectra of 4L MoSe$_2$ crystals of different stacking orientations extracted

In Figure S4, high-frequency Raman spectra for different stacking configurations are shown for the same regions for which each low-frequency spectrum was extracted. These spectra have been normalized to the $A'_1/A_{1g}$ peak and offset for clarity. In Figure S4(a), the 2L-2H, -3R and -3R* spectra are shown, with no significant change in the spectra discernible between different stacking configurations. In Figure S4(b), the 3L spectra for differing stacking configurations are shown, with no major change in the spectra between different stacking configurations, however a small relative intensity change in $B^1_{2g}$ mode can be observed in some of the 3L-3R-3R configurations. In Figure S7(c), the 4L spectra for differing stacking configurations are shown, with no observable change in the spectra between stacking configurations.

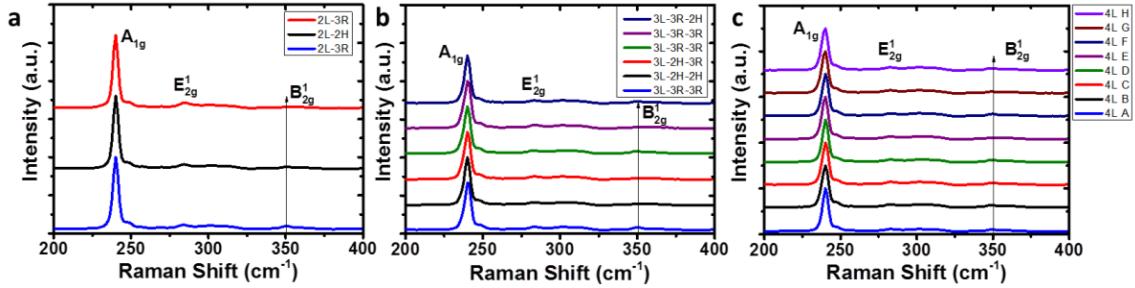

Figure S4 – Comparison of high-frequency Raman spectra for different stacking configurations within each layer number - (a) 2L high-frequency Raman spectra (b) 3L high-frequency Raman spectra (c) 4L high-frequency Raman spectra

**Independent Assessment of Layer number with AFM**

Much of the analysis in this work relies on the identification of layer number or stacking polytype for different TMDs using Raman spectroscopy. Atomic force microscopy (AFM) was used as an independent means of measuring the layer thickness, in order to verify the layer-number assignation from Raman spectroscopy and also to show that there was no change in thickness for different stacking polytypes of the same layer thickness. Figure S5 and S6 depict AFM analysis of two of the $MoSe_2$ regions analyzed in the main text. AFM measurements were carried out using a Veeco Dimension 3100 in tapping mode, with 40 N/m probes from Budget Sensors. In Figure S5(a) and (b), the maps of low-frequency maximum and position of maximum intensity have been reproduced from Figure 5 in the main text, along with the relevant optical image in Figure S5(c). Figure S5(d) shows an AFM image of this region with layer-thickness assignments from Raman spectroscopy overlaid; this has been rotated slightly in order to facilitate a direct comparison between AFM and Raman images. Figures S5(e-h) reproduce the low-frequency maps shown in Figure 5(e), (f), (h) and (i), in the main text, respectively. The centre triangle in Figure S5(g) is present due to overlap of peaks at this frequency for 3L with those of higher layer numbers. Figure S5(i) shows an AFM height map over this crystal region, with several height profiles marked. These correspond to height profiles 1-5 in Figure S5. Height profile 1 shows a decrease in layer number from 3L to 2L. Height profile 2 shows no change in layer number across a 3L region, which contains several different stacking polytypes according to low-frequency Raman analysis. Height profile 3 shows an increase in layer number from a 2L region to a 3L region. Height profile 4 shows an increase in layer number from 0L (background) to a 2L region. Height profile 5 shows no increase in height across 2L regions of different stacking polytype, as shown through low-frequency Raman analysis, and then an increase on going from a 2L region to a 3L region. The height differences between different layers in each instance are labelled on each profile, and range from 0.9-1 nm per layer, comparable to previous reports of CVD grown TMDs[8, 9]. Decorative contamination can be seen in the AFM scans, which has been observed previously for CVD grown TMDs[2, 10] – this is likely due to preferential particle adsorption at the growth front, or atmospheric contamination post growth. These height profiles support the layer numbers previously identified by Raman spectroscopic measurements and further show the lack of any measurable change in height across different stacking polytypes within a given layer number. This illustrates that low-frequency Raman spectroscopy can identify features which cannot be observed using AFM alone.

Figure S6 shows further analysis of an area of CVD MoSe$_2$ analysed in the main text. In Figure S6(a) and (b), the maps of low-frequency maximum and position of maximum intensity of the relevant crystal have been reproduced from Figure 5 in the main text, along with the relevant optical image in Figure S6(c). Figure S6(d) shows an AFM image of this region with layer-thickness assignations from Raman spectroscopy overlaid, which has been rotated slightly in order to facilitate a direct comparison of AFM and Raman images. Figures S6(e-h) reproduce the low-frequency maps shown in Figure 6(e), (f), (h) and (i) in the main text, respectively.

Figure S6(i) shows an AFM height map over this crystal region, with several height profiles marked. These correspond to height profiles 1-5 in Figure S6. Height profile 1 shows an increase from 0L to 3L. Height profile 2 shows an increase from 2L to 3L. Height profile 3 shows no change across regions spanning different 3L stacking polytypes. It should be noted here that some interesting corrugations are present on the surface crossing height profile 3 which merit further investigation. Height profile 4 shows an initial increase from 2L to 3L, followed by no change in layer number across regions consisting of different 3L stacking polytype. Height profile 5 shows an increase from 0L to 3L and back down to 0L. Height profile 6 shows an initial increase of 2L, followed by an increase from 2L to 3L. The height differences between different layers in each instance are labelled on each profile, and range from 0.9-1 nm per layer, comparable to previous reports of CVD grown TMDs[8, 9]. These height profiles support the layer numbers previously identified by Raman spectroscopic measurements, and further show the lack of any measurable change in height across different stacking polytypes within a given layer number.

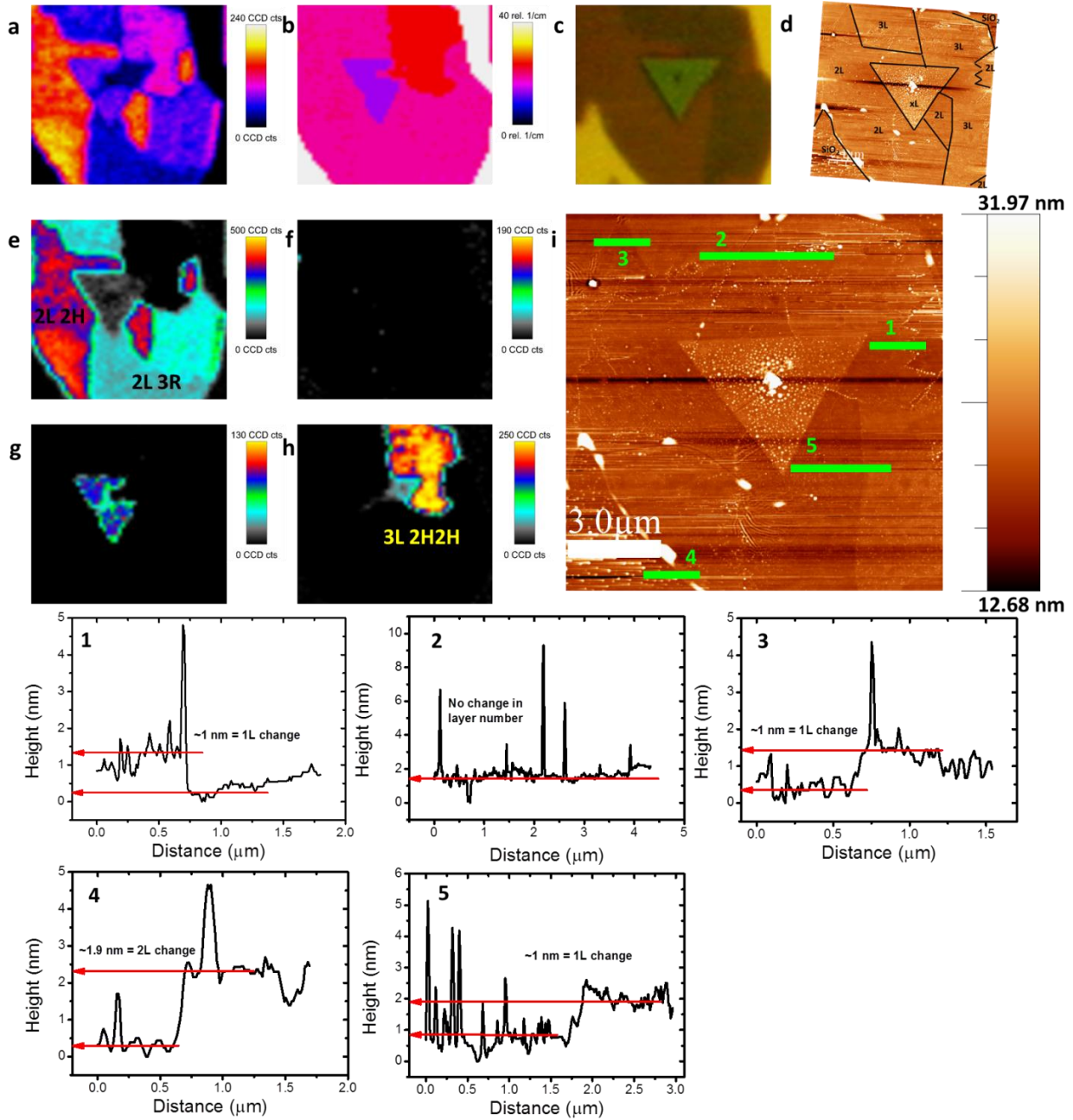

Figure S5 – (a) Peak intensity map over the range 10-50 cm$^{-1}$ (b) Map of position of maximum peak intensity of the low-frequency Raman modes in the range of 0-40 cm$^{-1}$ (c) Optical image of CVD MoSe$_2$ crystal from the area shown in Figures 4 and 5 in the main text (d) AFM scan with layer numbers overlaid as identified by Raman spectroscopy (e) Peak intensity map for 2L MoSe$_2$ SM at ~18 cm$^{-1}$ (f) Peak intensity map for 2L MoSe$_2$ LBM at ~29 cm$^{-1}$ (g) Peak intensity map for 3L MoSe$_2$ at ~13 cm$^{-1}$ (h) Peak intensity map for 3L MoSe$_2$ at ~24 cm$^{-1}$ (i) AFM height scan over MoSe$_2$ crystal with height profiles 1-5 labelled in overlay.

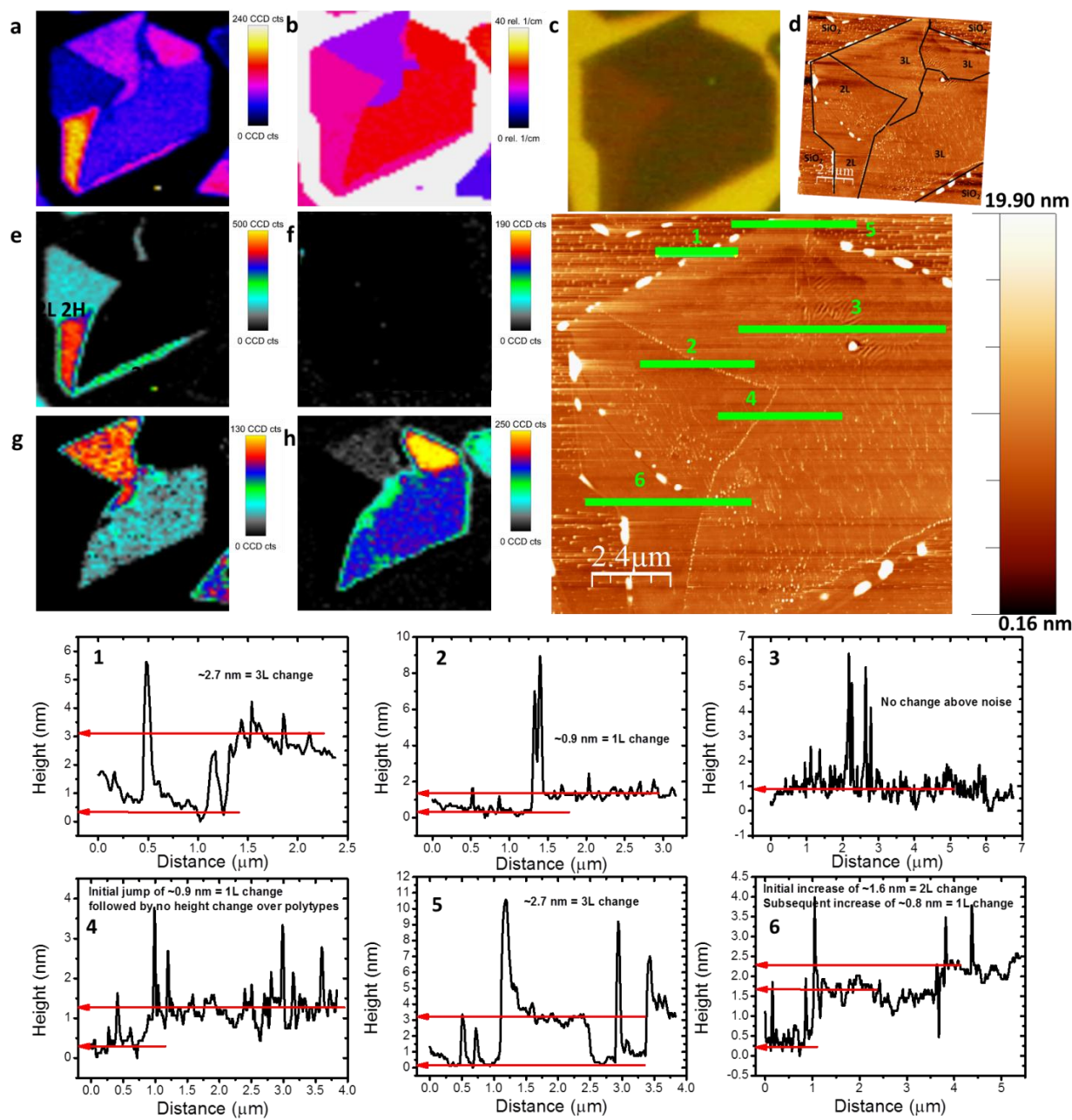

Figure S6 – (a) Peak intensity map over the range 10-50 cm$^{-1}$ (b) Map of position of maximum peak intensity of the low-frequency Raman modes in the range of 0-40 cm$^{-1}$ (c) Optical image of CVD MoSe$_2$ crystal from the area shown in Figures 4 and 5 in the main text (d) AFM scan with layer numbers overlaid as identified by Raman spectroscopy (e) Peak intensity map for 2L MoSe$_2$ SM at ~18 cm$^{-1}$ (f) Peak intensity map for 2L MoSe$_2$ LBM at ~29 cm$^{-1}$ (g) Peak intensity map for 3L MoSe$_2$ at ~13 cm$^{-1}$ (h) Peak intensity map for 3L MoSe$_2$ at ~24 cm$^{-1}$ (i) AFM height scan over MoSe$_2$ crystal with height profiles 1-6 labelled in overlay.

**Additional WSe$_2$ Analysis**

The peak position map for the $A'_1/A_{1g}$ and $E'/E_g/E^1_{2g}$ overlapping mode is shown in Figure S7(a), with no discernible differences noted between different layer numbers. It is clear from correlation of Figure 6(b) in the main text with Figure S7(a) here that this peak maximum red-shifts to lower wavenumbers as layer number increases from 1 to 2+ layers, before blue-shifting to higher wavenumbers as layer number approaches bulk values. This is attributed to the enhancement of $E'/E_g/E^1_{2g}$ mode as layer number increases from 1 to 2+ layers, followed by the domination of the out-of-plane $A'_1/A_{1g}$ mode as more layer numbers contribute in bulk-like regions, as is indicated in the spectra in Figure 6(d) of the main text. The *2LA(M)* mode peak position map is shown in Figure S7(b), with a notable blue shift between mono- and few-layer samples – however, this may simply be due to the enhanced relative intensity of the *2LA(M)* mode to the $A'_1/A_{1g}$ and $E'/E_g$ overlapping mode with increasing layer number. The PL position for monolayer appears at a wavelength of ~765 nm consistent with those previously reported for CVD-grown[11, 12] and mechanically exfoliated WSe$_2$[7, 13], showing an expected increase in wavelength with increasing layer number.

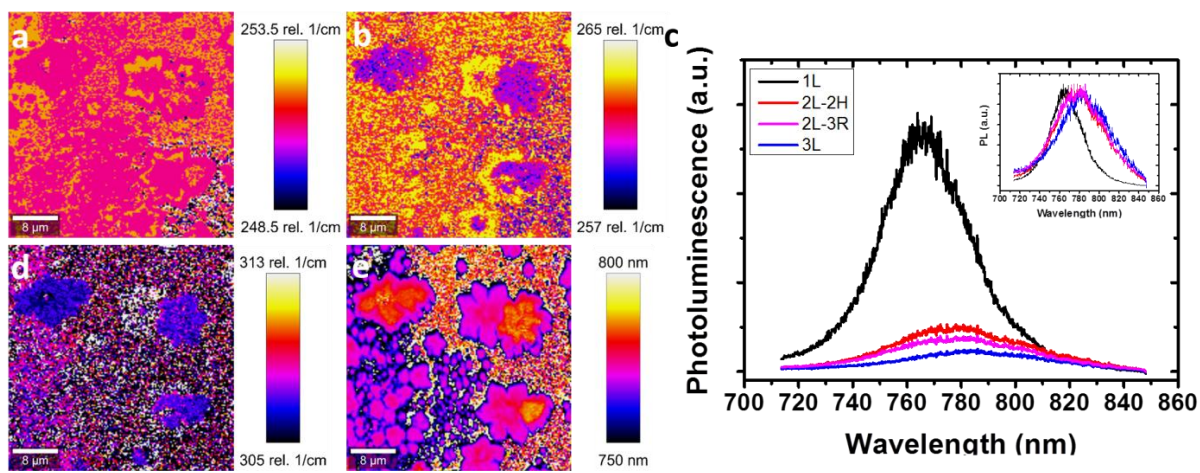

Figure S7 - Additional Raman and PL mapping of $WSe_2$ crystals of different layer thickness extracted from the regions shown in Figure 6(a) of the main text – (a) Position map of $A'_1/A_{1g}$ and $E'/E_g/E^1_{2g}$ mode overlapping peak (b) Position map of $2LA(M)$ mode (c) PL spectra of $WSe_2$ crystals of different layer thickness extracted from the regions shown in Figure 6(a) of the main text. Inset: Normalized version of the same spectra highlighting the evolution in peak position with changing layer thickness. (d) Position map of maximum $B^1_{2g}$ mode (e) PL position map.

### Additional $WS_2$ Analysis

A map of peak position for the combination mode centred at ~352 cm$^{-1}$ is shown in Figure S8(a). This peak is a combination of contributions from the $2LA(M)$ and $E'/E_g/E^1_{2g}$ modes that coincidentally overlap at this Raman shift. A red shift in position of maximum intensity with increasing layer number is evident, likely due to enhanced contributions with increasing layer number from the $E'/E_g/E^1_{2g}$ mode at a higher Raman shift than $2LA(M)$. A position maximum map of the $A'_1/A_{1g}$ mode intensity maximum is shown in Figure S8(b). A blue shift in position of maximum intensity is clear as layer number increases. This is due to stronger interlayer contributions to the phonon restoring forces as layer number increases, resulting in a stiffening of the out-of-plane $A'_1/A_{1g}$ mode, as the vibrations of this mode are more strongly affected by forces between the layers[14]. A peak intensity map of the zone-edge $LA(M)$ phonon[14] is shown in Figure S8(c), with the corresponding spectra shown in Figure S8(h), which have been normalized to the $2LA(M)$ mode and offset for clarity. It is clear that this $LA(M)$ mode has strongest intensity in monolayer crystals, with a negligible change in position with increasing layer number. The behaviour of intensity with layer number in *LA(M)* can be compared to that of the low-frequency resonance peak discussed for $WS_2$ in the main text, and can be compared directly on the spectra here.

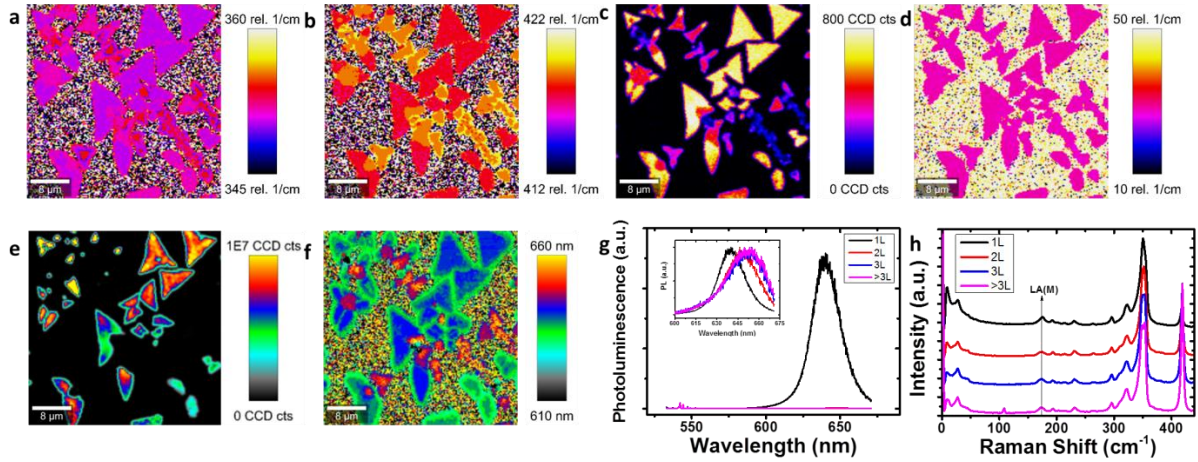

Figure S8 - Additional Raman and PL mapping of $WS_2$ crystals of different layer thickness extracted from the regions shown in the optical image in Figure 8(a) of the main text - (a) Position map of *2LA(M)* mode and $E'/E_g/E^1_{2g}$ mode overlap (b) Map of $A'_1/A_{1g}$ mode position (c) Map of *LA(M)* mode intensity (d) Map of position of low-frequency mode at ~27 cm$^{-1}$ (e) PL intensity map (f) Position map of PL maxima (g) Corresponding PL spectra of 1, 2, 3 and 3+L $WS_2$. Inset shows plots normalized to PL intensity (h) Spectra of 1, 2, 3 and 3+L $WS_2$ centred on the *LA(M)* mode.

**Growth Schematic**

The schematic in Figure S9(a) shows the quartz tube furnace configuration used for fabrication of CVD materials. A 10% $H_2$/Ar flow enters through the gas inlet as labelled. In Zone 2, sulfur or selenium solid precursors are placed, and evaporated at temperatures indicated. This vapour then flows downstream to microreactors placed in Zone 1, before being evacuated through the pump. A microreactor schematic is shown in Figure S9(b), as reported previously[2].

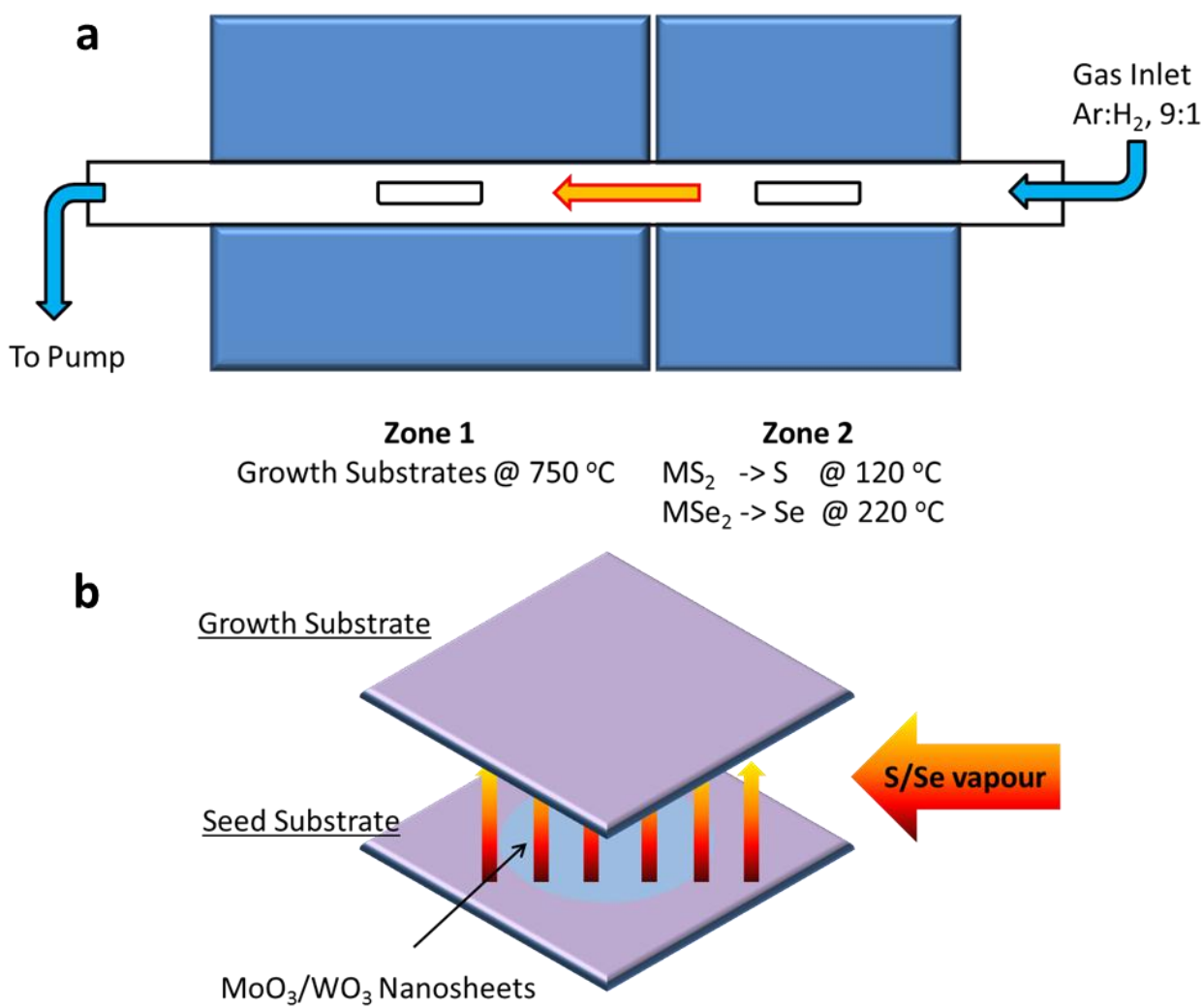

Figure S9 – (a) Schematic of furnace setup. Chalcogen powder is melted downstream and flowed through the microreactor (b) Schematic of CVD microreactor formed between the seed and target substrates, where sulfur reacts with $MO_3$ nanosheets to form $MX_2$ layers on the top substrate.